\documentclass[pre,superscriptaddress,showpacs,floatfix]{revtex4}

\usepackage{graphicx}
\usepackage{epsfig}
\usepackage[english]{babel}
\usepackage{bm,amsmath,amssymb}

\newcommand{\si}{\sin{\theta}}
\newcommand{\co}{\cos{\theta}}

\newcommand{\e}{\varepsilon}

\newcommand{\sit}{\sin^2{\theta}}

\newcommand{\cotg}{\cot{\theta}}

\newcommand{\sinul}{\sin{\theta_0}}
\newcommand{\sitnul}{\sin^2{\theta_0}}
\newcommand{\sidnul}{\sin^3{\theta_0}}
\newcommand{\siviernul}{\sin^4{\theta_0}}

\newcommand{\su}{\sum_iS_i\rho_i}

\newcommand{\Te}{T_{\text{eff}}}

\newcommand{\xgem}{\langle x\rangle}
\newcommand{\xtgem}{\langle x^2\rangle}
\newcommand{\xdgem}{\langle x^3\rangle}
\newcommand{\xviergem}{\langle x^4\rangle}
\newcommand{\xvijfgem}{\langle x^5\rangle}
\newcommand{\xzesgem}{\langle x^6\rangle}

\newcommand{\xgems}{\langle x\rangle}
\newcommand{\xtgems}{\langle x^2\rangle}

\newcommand{\Vgem}{\langle V\rangle}
\newcommand{\Vgems}{\langle V\rangle}

\newcommand{\fhalf}{\ f[1/2]\ }
\newcommand{\fdrietw}{\ f[3/2]\ }
\newcommand{\fminhalf}{\ f[-1/2]\ }
\newcommand{\fmindrietw}{\ f[-3/2]\ }
\newcommand{\fvijftw}{\ f[5/2]\ }
\newcommand{\gnul}{\ g[0]\ }
\newcommand{\gtwee}{\ g[2]\ }

\newcommand{\hhalf}{h[1/2]}
\newcommand{\hdrietw}{h[3/2]}
\newcommand{\hminhalf}{h[-1/2]}
\newcommand{\hnul}{h[0]}
\newcommand{\heen}{h[1]}
\newcommand{\htwee}{h[2]}

\newcommand{\sigem}{\langle\sin{\theta}\rangle}

\newcommand{\sitgem}{\langle\sin^2{\theta}\rangle}
\newcommand{\sidgem}{\langle\sin^3{\theta}\rangle}

\newcommand{\sitgemi}{\langle\sin^2{\theta}\rangle_{_i}}
\newcommand{\sidgemi}{\langle\sin^3{\theta}\rangle_{_i}}
\newcommand{\siviergemi}{\langle\sin^4{\theta}\rangle_{_i}}
\newcommand{\sivijfgemi}{\langle\sin^5{\theta}\rangle_{_i}}
\newcommand{\sizesgemi}{\langle\sin^6{\theta}\rangle_{_i}}
\newcommand{\sizevengemi}{\langle\sin^7{\theta}\rangle_{_i}}

\begin{document}
\title{Rectification of thermal fluctuations in ideal gases}

\author{P. Meurs}
\affiliation{Limburgs Universitair Centrum, B-3590 Diepenbeek,
Belgium}

\author{C. Van den Broeck}
\affiliation{Limburgs Universitair Centrum, B-3590 Diepenbeek,
Belgium}

\author{A. Garcia}
\affiliation{Department of Physics, San Jose State University, San
Jose, CA 95192-0106}

\date{\today}

\begin{abstract}
We calculate the systematic average speed of the adiabatic piston
and a thermal Brownian motor, introduced in [Van den Broeck, Kawai
and Meurs, \emph{Microscopic analysis of a thermal Brownian
motor}, to appear in Phys. Rev. Lett.], by an expansion of the
Boltzmann equation and compare with the exact numerical solution.
\end{abstract}


\maketitle

\section{Introduction}

Can thermal fluctuations be rectified? Ever since Maxwell
\cite{maxwell} raised this question with his famous thought
experiment involving a so-called Maxwell demon, it has been the
object of debate in both the thermodynamics and statistical
physics community. The mainstream opinion is that rectification is
impossible in a system at equilibrium. Indeed the property of
detailed balance, which was discovered by Onsager \cite{onsager},
and which turns out to be a basic characteristic of the steady
state distribution in area preserving time-reversible dynamical
systems (in particular Hamiltonian systems) \cite{skordos}, states
that any transition between two states (defined as regions in
phase space of nonzero measure and even in the speed) occurs as
frequently as the time-reversed transition. The separate issue,
first introduced by Szilard \cite{szilard},  of involving an
``intelligent observer'' that tracks the direction of these
transitions, making possible the rectification by interventions at
the right moment, has a contorted history of its own. It turns out
that the engendered rectification is offset by the entropic cost
of processing (and more precisely of erasing) the  involved
information \cite{landauer}. Another more recent
debate involving entangled quantum systems \cite{nieuwenhuizen} is still ongoing.\\
Apart from the fundamental interest in the subject, a number of
recent developments have put the issue of rectifying thermal
fluctuations back on the agenda. First, we mention the observation
that thermal fluctuations can in principle be rectified if the
system under consideration operates under nonequilibrium
conditions. The past decade has witnessed a surge in the
literature on the subject  of the so-called Brownian motors
\cite{reimann}. Such motors possibly explain, amongst other,
phenomena such as transport and force generation in biological
systems. Second, our ability to observe, manipulate or even
fabricate objects on the nanoscale prompts us to look into new
procedures to regulate such small systems, possibly by
exploiting the effects of thermal fluctuations in a constructive way.\\
Even though several constructions have been envisaged to discuss
the issue of rectification in more detail, including for example
the Smoluchowski-Feynman ratchet \cite{feynman}, the issue of
thermal fluctuations in a system with nonlinear friction
\cite{alkemade} and the thermal diode \cite{vankampen, sokolov},
no exactly solvable model has been put forward. In this paper, we
will present two fully microscopic Hamiltonian models, in which
the rectification of thermal fluctuations can be studied in
analytic detail. Versions of the first model have appeared in the
literature for some time under the name of Rayleigh piston
\cite{alkemade} or adiabatic piston \cite{adpiston, callen,
gruber}, see also \cite{handrich}. The second model, to which we
will refer as thermal Brownian motor, was introduced in a recent
paper \cite{vandenbroeckprl}. Both models involve a small object
simultaneously in contact with two infinite reservoirs of ideal
gases, each separately at equilibrium but possibly at a different
temperature and/or density. A Boltzmann-Master equation provides a
microscopically exact starting point to study the motion of the
object. As a result of the rectification of the nonequilibrium
fluctuations, the object acquires a systematic average speed,
which will be calculated exactly via a perturbative solution of
the Boltzmann equation, with the ratio of the mass of the gas
particles over that of the object as the
small parameter.\\
The organization of this paper is as follows.  We start in section
\ref{s1} by reviewing the general framework and the type of
construction for the Brownian motor that we have in mind. The main
technical ingredients are closely related to the so-called
$1/\Omega$-expansion of van Kampen \cite{vankampen}. In section
\ref{s2} we turn to a detailed presentation and discussion of the
adiabatic piston. The rectification has in this case been
investigated to lowest order by Gruber and Piasecki \cite{gruber}.
We present a streamlined derivation allowing to go two orders
further in the expansion. Next, in section \ref{s3}, we discuss
the more surprising thermal Brownian motor in which the motion
derives from the spatial asymmetry of the object itself
\cite{vandenbroeckprl}. We again calculate the three first
relevant terms in the expansion of the average speed. Finally, in
section \ref{s4}, the obtained analytic results are compared with
a direct numerical solution of the Boltzmann-Master equation and
with previous molecular dynamics simulations
\cite{vandenbroeckprl}.

\section{Expansion of the Boltzmann-Master Equation}\label{s1}

Consider a closed, convex and rigid object with a single degree of
freedom, moving in a gas. To obtain a microscopically exact
equation for the speed $V$ of this object, we will consider the ideal gas limit in which: \\
(1) the gas particles undergo
instantaneous and perfectly elastic collisions with the object,\\
(2) the mean free
path of the particles is much larger than the linear dimensions of the object (regime of a large Knudsen number),\\
(3) the (ideal) gas is initially at equilibrium, and hence at all times: the perturbation due to the
collisions with  the object are negligible in an infinitely large reservoir.\\
With these assumptions, there are no precollisional correlations
between the speed of the object and those of the impinging gas
particles, hence the Boltzmann ansatz of molecular chaos is exact
\cite{dorfman}. In fact, since the  collisions with the gas
particles occur at random and uncorrelated in time, the speed $V$
of the object is a Markov process and its probability density
obeys a Boltzmann-Master equation of the following form:
\begin{eqnarray}
\frac{\partial P(V,t)}{\partial t}=\int{dV'
\Big[W(V|V')P(V',t)-W(V'|V)P(V,t)\Big].}\label{mastereq}
\end{eqnarray}
$W(V|V')$ represents the transition probability per unit time to change the speed
of the object from $V'$ to $V$. Its detailed form can be easily obtained following arguments familiar from
the kinetic theory of gases.\\
To construct a model for a Brownian motor, two additional
ingredients need to be introduced. First we have to operate under
nonequilibrium conditions. This can most easily be achieved by
considering that the object interacts not with a single but with
two  ideal gases, both at equilibrium in a separate reservoir,
each at its own temperature and density. The physical separation
(no particle exchange) between the gases  can be achieved by using
the object itself as a barrier (adiabatic piston) or by assuming
that the object consists out of two rigidly linked  (closed and
convex) units, each moving in one of the separate reservoirs
containing the gases. Second, we need to break the spatial
symmetry. In the adiabatic piston this is achieved by the
asymmetric distribution of the gases with respect to the piston.
In the thermal Brownian motor, at least one of the constitutive
units needs to be spatially asymmetric. With these modifications
in mind, we can still conclude that Eq. (\ref{mastereq}) remains
valid, but the transition probability is now a sum of the
contributions representing the collisions with the
particles of each gas.\\
With the ingredients for a Brownian motor thus available, we
expect that the object can rectify the fluctuating force resulting
from the collisions with the gas particles. Hence it will develop
a steady state average nonzero systematic speed, which we set out
to calculate analytically.  Unfortunately, an explicit exact
solution of Eq. (\ref{mastereq}) cannot be obtained  even at the
steady state, and  a perturbative solution is required. Since we
expect that the rectification disappears in the limit of a
macroscopic object, a natural expansion parameter is the ratio of
the mass $m$ of the gas particle over the mass $M$ of the object.
More precisely, we will use $\e=\sqrt{m/M}$ as the expansion
parameter. In fact this type of expansion is very familiar for the
equilibrium version of the adiabatic piston, namely the so-called
Rayleigh particle. It has been developed with the primary aim of
deriving exact Langevin equations from microscopic theory and
culminated in the more general well-known $1/\Omega$ expansion of
van Kampen \cite{vankampen}. With the aim of streamlining this
procedure for the direct calculation of the average drift
velocity, with special attention to higher order corrections, we
briefly review the technical details. First it is advantageous to
introduce the  transition probability $W(V';r)=W(V|V')$, defined
in terms of the jump amplitude $r=V-V'$, since the latter jumps
are anticipated to become small in the limit $\e \rightarrow 0$.
One can then rewrite the Master equation as follows:
\begin{eqnarray}\label{masterjump}
\frac{\partial P(V,t)}{\partial
t}&=&\int{W(V-r;r)P(V-r,t)dr}-P(V,t)\int{W(V;-r)dr}.
\end{eqnarray}
A Taylor expansion of the transition probability in the first
integral of Eq. (\ref{masterjump}) with respect to the jump
amplitude leads to an equivalent expression
 under the form  of the Kramers-Moyal expansion:
\begin{eqnarray}
\frac{\partial P(V,t)}{\partial
t}=\sum_{n=1}^{\infty}{\frac{(-1)^n}{n!}
\left(\frac{d}{dV}\right)^n\left\{a_n(V)P(V,t)\right\},}\label{kramers-moyal}
\end{eqnarray}
with the so-called ``jump moments'' given by
\begin{eqnarray}\label{jumpmomentGeneral}
a_n(V)=\int{r^n W(V;r)dr}.
\end{eqnarray}

Since the change in the speed of our object of mass $M$, i.e., the
jump amplitude $r$, will, upon colliding with a particle of mass
$m$, be of order $\e^2=m/M$, the Kramers-Moyal expansion appears
to provide the requested expansion in our small parameter.
However, the parameter $M$ will also appear implicitly in the
speed $V$. Indeed, we expect that the object will, in the
stationary regime, exhibit thermal fluctuations at an effective
temperature $\Te$, i.e., $\frac{1}{2}M\langle
V^2\rangle=\frac{1}{2}k_B\Te$. To take this into account, we
switch to a dimensionless variable $x$ of order 1:
\begin{eqnarray}\label{defx}
x&=&\sqrt{\frac{M}{k_B \Te}}V.
\end{eqnarray}
The explicit value of $\Te$ will be determined below by
self-consistency, more precisely from the condition  $\langle
x^2\rangle=1$ to first order in $\e$. The probability density
$P(x,t)$ for the  new variable $x$ thus obeys the following
equation:
\begin{eqnarray}
\frac{\partial P(x,t)}{\partial
t}=\sum_{n=1}^{\infty}{\frac{(-1)^n}{n!}
\left(\frac{d}{dx}\right)^n\left\{A_n(x)P(x,t)\right\},}\label{kramers-moyal}
\end{eqnarray}
with rescaled jump moments, $A_n(x)$, defined as
\begin{eqnarray}\label{defAn}
A_n(x)&=&\left(\sqrt{\frac{M}{k_B \Te}}\right)^n
a_n(x).
\end{eqnarray}
Equivalently, and of more interest to us,
the following set of  coupled equations determine the moments
$\langle x^n\rangle=\int x^n P(x,t) dx$:
\begin{eqnarray}\label{momentvglx}
\partial_t\langle x\rangle\;&=&\langle A_1(x)\rangle\nonumber\\
\partial_t \langle x^2\rangle&=&2\langle x A_1(x)\rangle + \langle A_2(x)\rangle\nonumber\\
\partial_t \langle x^3\rangle &=&3\langle x^2 A_1(x)\rangle
+3\langle x A_2(x)\rangle + \langle A_3(x)\rangle\nonumber\\
\partial_t \langle x^4\rangle &=&4\langle x^3 A_1(x)\rangle
+6\langle x^2 A_2(x)\rangle + 4\langle x A_3(x)\rangle +\langle
A_4(x)\rangle\nonumber\\
\partial_t \langle x^5\rangle &=&5\langle x^4 A_1(x)\rangle
+10\langle x^3 A_2(x)\rangle + 10\langle x^2 A_3(x)\rangle + 5
\langle x A_4(x)\rangle +\langle A_5(x)\rangle\nonumber\\
\partial_t \langle x^6\rangle &=&6\langle x^5 A_1(x)\rangle
+15\langle x^4 A_2(x)\rangle + 20\langle x^3 A_3(x)\rangle + 15
\langle x^2 A_4(x)\rangle +6 \langle x A_5(x)\rangle+\langle A_6(x)\rangle\nonumber\\
&\cdots&
\end{eqnarray}

The exact solution of this coupled set of equations is as hopeless
and equally difficult as the full Boltzmann-Master equation.
However, a Taylor expansion in $\e$ shows that the equations are
no longer fully coupled and the calculation of a moment up to a
finite order reduces to an in principle simple (but in practice
tedious) algebraic problem.

\section{The Adiabatic Piston}\label{s2}

\subsection{Motivation}

\begin{figure}[bt]
\begin{center}
\includegraphics[width=3in]{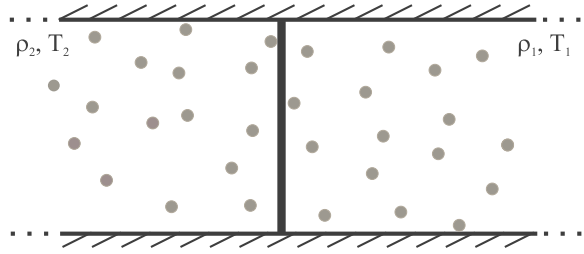}
\caption{The adiabatic piston.} \label{fig:adiabatic}
\end{center}
\end{figure}

In Fig. \ref{fig:adiabatic}, we have represented in a schematic
way the construction of the Rayleigh piston and of its
nonequilibrium version known as the adiabatic piston. We
concentrate here on an (infinite) two-dimensional system, for
reasons of simplicity. The piston is considered to be a single
``flat'' particle of length $L$ and mass $M$ with a unique degree
of freedom, namely its position $x$ along the horizontal axis.
Since the piston has no internal degrees of freedom, it can not
transfer energy by ``hidden'' microscopic degrees of freedom. The
absence of a corresponding heat exchange prompted the use of the
name ``adiabatic piston''. The piston is moving inside an infinite
rectangle separating the gases to its right and left from each
other. These gases are initially taken separately in equilibrium,
but not necessarily at equilibrium with each other. In the
thermodynamic context of a macroscopic piston, this construction
is an example of an indeterminate problem, i.e., the final
position of the piston can not be predicted by the criterion of
maximizing the total entropy, since it  depends on the initial
preparation of the gases \cite{callen}, see also \cite{gruber}.
The case of interest to us is when the mass of the piston is not
macroscopically large, i.e., finite $\e=\sqrt{m/M}$. When
operating furthermore at equilibrium, this Rayleigh piston
provides  an exactly solvable model, allowing for example, the
rigorous derivation of a linear Langevin equation appearing as the
first nontrivial limit of the Boltzmann-Master equation in the
limit $\e\rightarrow 0$. When the left and right gases are not at
equilibrium, but exert equal pressure on the piston, the model
becomes an example of a Brownian motor, which is able to perform
work by rectifying pressure fluctuations \cite{gruber}. In doing
so, the single degree of freedom $x$ also plays the role of a
microscopic thermal conductivity, an issue that is quite relevant
to other models of Brownian motors \cite{parrondo}. Since this
model is essentially one-dimensional and the related calculations
are relatively simple, we include it in this paper to illustrate
the calculation procedure and at the same time to derive novel
results for the average drift speed up to order $\e^5$.

\subsection{Presentation of the model}

 The ideal gases in the right and left
compartments, separated from each other by the piston, are each at
equilibrium with Maxwellian velocity distributions at temperatures
$T_1$ and $T_2$, and with uniform particle densities $\rho_1$ and
$\rho_2$, respectively.
Since we are mainly interested in the rectification of fluctuations, we will focus on  the case of mechanical
equilibrium with equal pressure on both sides of the piston, i.e.,
$\rho_1T_1=\rho_2T_2$. \\
The motion of the piston is determined by the laws of Newton.
Hence its velocity only changes, say from  $V'$ to $V$, when it
undergoes a collision with a gas particle, its ($x$-component of
the) velocity going from $v'_x$ to $v_x$. Conservation of energy
and momentum determines the post-collisional speeds in terms of
the pre-collisional ones:
\begin{eqnarray*}
\frac{1}{2} mv_x^2 + \frac{1}{2} MV^2 &=& \frac{1}{2} mv'_x{^2} +
\frac{1}{2} MV'{^2}\\
mv_x + MV &=& mv'_{x} + MV',
\end{eqnarray*}
implying:
\begin{eqnarray}
V=V'+\frac{2m}{m+M}(v'_x-V').
\end{eqnarray}
The transition probability $W(V|V')$ then follows from standard
arguments in kinetic theory of gases: one evaluates the frequency
of collisions of gas particles of a given speed and subsequently
integrates over all the speeds. Note that we have two separate
contributions from the gas  right ($\rho_1$ and $T_1$) and left
($\rho_2$ and $T_2$). The result reads:
\begin{eqnarray*}
W(V|V')=\left\{\begin{array}{ll}
 L\rho_1 \int_{-\infty}^{+\infty}{dv_x} (v_x-V') H\left[v_x-V'\right]
\phi_1(v_x)\delta\left[V'+\frac{2m}{m+M}(v_x-V')-V\right] & \textrm{ if }V<V'\nonumber\\
 L \rho_2 \int_{-\infty}^{+\infty}{dv_x} (V'-v_x) H\left[V'-v_x\right]
\phi_2(v_x)\delta\left[V'+\frac{2m}{m+M}(v_x-V')-V\right] &
\textrm{ if }V>V',
\end{array}\right.
\label{tranprobad1}
\end{eqnarray*}
with $H$ the Heaviside function, $\delta$ the Dirac distribution
and $\phi_i$ the Maxwell-Boltzmann distribution at temperature
$T_i$:
\begin{eqnarray*}
\phi_i(v_x)=\sqrt{\frac{m}{2 \pi k_B
T_i}}\exp{\left(\frac{-mv_x^2}{2 k_B T_i}\right).}
\end{eqnarray*}
Performing the integrals over the speed gives the following explicit
result for the transition probability:
\begin{eqnarray}
W(V|V')&=&L\rho_1 \left[\frac{m+M}{2m}\right]^2 (V'-V)
H\left[V'-V\right]
\phi_1(V'-\frac{m+M}{2m}(V'-V))\nonumber\\
&+& L \rho_2 \left[\frac{m+M}{2m}\right]^2 (V-V')
H\left[V-V'\right] \phi_2(V'-\frac{m+M}{2m}(V'-V)).
\label{tranprobad2}
\end{eqnarray}
From the transition probability, the rescaled jump moments
$A_n(x)$ (\ref{defAn}) can be calculated. The exact expression for
the $n$-th jump moment is as follows :
\begin{eqnarray}\label{adAn}
A_n(x)&=&2^{(3n-1)/2}L\sqrt{\frac{k_B}{\pi
m}}\frac{\e^n}{(1+\e^2)^{n}} \Te^{-n/2}
\Gamma\left[\frac{2+n}{2}\right] \left((-1)^n \rho_1
T_1^{(n+1)/2}\exp\left[-\frac{\Te}{2T_1}x^2\e^2\right]
\Phi\left[\frac{2+n}{2},\frac{1}{2},\frac{\Te}{2 T_1}x^2\e^2\right]\right.\nonumber\\
& &\ \ \ \ \ \ \ \ \ \ \ +\left.\rho_2 T_2^{(n+1)/2}\exp\left[-\frac{\Te}{2T_2}x^2\e^2\right]
\Phi\left[\frac{2+n}{2},\frac{1}{2},\frac{\Te}{2 T_2}x^2\e^2\right]\right)\nonumber\\
&+&2^{3n/2}L\sqrt{\frac{k_B}{\pi
m}}\frac{\e^{n+1}}{(1+\e^2)^{n}}\Te^{(1-n)/2}
\Gamma\left[\frac{3+n}{2}\right]x \left((-1)^n \rho_1
T_1^{n/2}\exp\left[-\frac{\Te}{2T_1}x^2\e^2\right]
\Phi\left[\frac{3+n}{2},\frac{3}{2},\frac{\Te}{2 T_1}x^2\e^2\right]\right.\nonumber\\
& & \ \ \ \ \ \ \ \ \ \ \ -\left.\rho_2
T_2^{n/2}\exp\left[-\frac{\Te}{2T_2}x^2\e^2\right]
\Phi\left[\frac{3+n}{2},\frac{3}{2},\frac{\Te}{2
T_2}x^2\e^2\right]\right), \label{jmap}
\end{eqnarray}
with $\Gamma$ the Gamma function:
\begin{eqnarray}\label{gamma}
\Gamma\left[1+k\right]&=&k!\nonumber\\
\Gamma\left[1+\frac{2k+1}{2}\right]&=&\frac{\sqrt{\pi}}{2^{k+1}}1\cdot3\cdot5\cdot\ldots\cdot(2k+1),
\end{eqnarray}
and
$\Phi$ the Kummer function, in its integral representation given by
\begin{eqnarray}\label{kummer}
\Phi[a,b,z]=\frac{\Gamma[b]}{\Gamma[b-a]\Gamma[a]}\int_0^1{e^{zt}t^{a-1}(1-t)^{b-a-1}dt}.
\end{eqnarray}

\subsection{Stationary speed}

\begin{figure}
\begin{center}
\includegraphics[width=7cm]{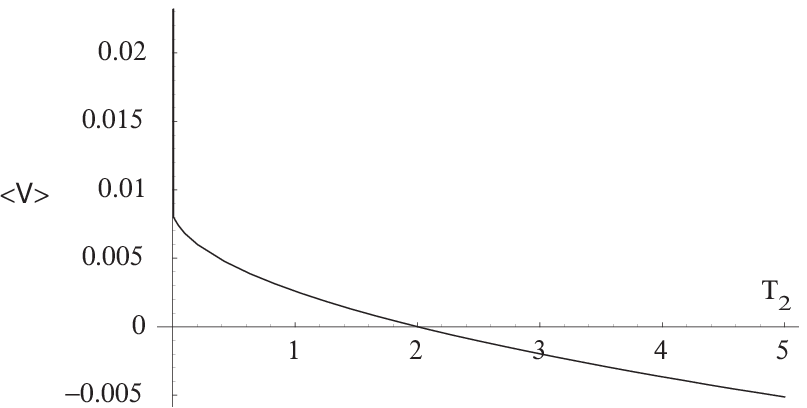}\hspace{2cm}
\includegraphics[width=7cm]{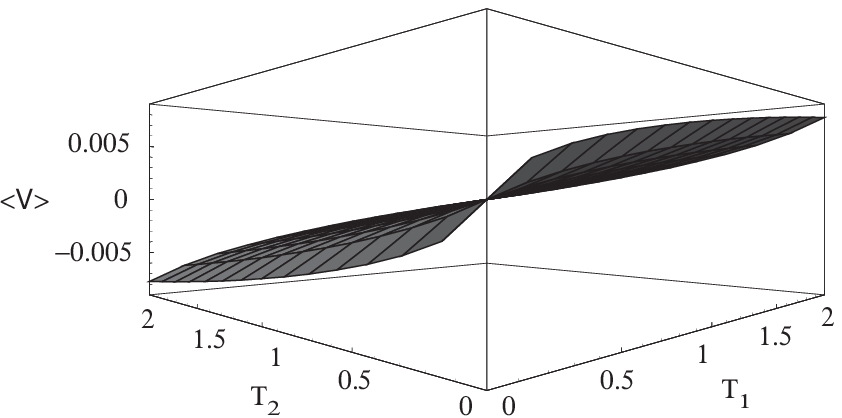}
\end{center}
\caption{Stationary average speed of the adiabatic piston
according to Eq. (\ref{adV5}). Left: For $\rho_1=0.25$ and
$T_1=2.0$ fixed, the speed is shown as a function of $T_2$. Note
that $\rho_2$ is determined by the condition of mechanical
equilibrium ($\rho_1 T_1=\rho_2 T_2$) and that the piston always
moves to the higher temperature region. Right: the speed of the
piston as a function of $T_1$ and $T_2$, for $\rho_1=0.25$ and
$\rho_2=\rho_1 T_1/T_2$. The velocity vanishes when $T_1=T_2$ and
is maximal for a large temperature difference. The following
parameter values were used: mass of the gas particles $m=1$, mass
of the piston $M=100$ and $k_B=1$ by choice of units.}
\label{fig:speedAd}
\end{figure}

The  moment equations (\ref{momentvglx}) form together with the
explicit expressions (\ref{jmap}) for the jump moments the
starting point for a straightforward perturbation in terms of the
small parameter $\e$. To simplify notation, we introduce:
\begin{eqnarray}
f(n)&=&L\sqrt{\frac{2}{\pi}}\sqrt{\frac{k_B}{m}}
\frac{\rho_1T_1^{n}+\rho_2T_2^n}{\Te^{n-\frac{1}{2}}}\\
g(n)&=&L\sqrt{\frac{k_B}{m}}\frac{\rho_1T_1^{n}-\rho_2T_2^n}{\Te^{n-\frac{1}{2}}}.
\end{eqnarray}
Also, the limit $\e \rightarrow 0$ entails a slowing down of the
motion of the piston, which can be accounted for by introducing a
new scaled time variable:
\begin{eqnarray*}
\tau=\e^2 t.
\end{eqnarray*}
The equations for the first and second moment, expanded up to
order $\e^5$ and $\e^4$ respectively, are as follows (the
expansion for higher moments up to the sixth moment can be found
in appendix A):
\begin{eqnarray}
\partial_{\tau}\xgem&=&-2\fhalf\xgem -\gnul\xtgem \e +
\frac{1}{3}\left(6 \fhalf \xgem - \fminhalf\xdgem\right)
\e^2 \label{admomentvgl1}\\
& &+\gnul\xtgem
\e^3+\left(-2\fhalf\xgem+\frac{1}{3}\fminhalf\xdgem
+\frac{1}{60}\fmindrietw\xvijfgem \right)\e^4\nonumber -
\gnul\xtgem \e^5+O(\e^6)\nonumber
\\
%
\partial_{\tau}\xtgem&=&4\left(\fdrietw  -
\fhalf\xtgem\right)-2\gnul\xdgem\e
+2\left(-4\fdrietw+5\fhalf\xtgem-\frac{1}{3}
\fminhalf\xviergem\right)\e^2\label{admomentvgl2}\\
& & +4\gnul\xdgem\e^3+\left(12\fdrietw-16\fhalf\xtgem
+\frac{7}{6}\fminhalf\xviergem+\frac{1}{30}\fmindrietw\xzesgem\right)\e^4+O(\e^5).\nonumber
\end{eqnarray}
Note that the condition of macroscopic equilibrium ($\rho_1
T_1=\rho_2T_2$) was used to derive these equations. In particular,
without this constraint, an additional term corresponding to a
constant, velocity-independent, force acting on the piston, would
be present in Eq. (\ref{admomentvgl1}).
\newline
To lowest order in $\e$, the equation for the first moment, Eq.
(\ref{admomentvgl1}), is not coupled to higher order moments. It
displays the usual linear relaxation term of the velocity, namely,
in original variables, $M\partial_t \langle V \rangle=-\gamma
\langle V \rangle$, with friction coefficient $\gamma$:
\begin{eqnarray}\label{adgamma}
\gamma=4L\sqrt{\frac{k_B
m}{2\pi}}\left(\rho_1\sqrt{T_1}+\rho_2\sqrt{T_2}\right).
\end{eqnarray}
For $T_1=T_2$ and (consequently) $\rho_1=\rho_2$, this result is in agreement with \cite{vankampen}.
We conclude that at this order of the perturbation, the steady state speed is zero. This is not surprising
since  any asymmetry is buried at the level of linear response theory.
\newline
Going beyond the lowest order, one enters into the  domain where
fluctuations and nonlinearity are intertwined. The first moment is
now coupled to the higher order moments. Therefore, we focus on
the steady state speed reached by the piston in the long time
limit.  We will omit, for simplicity of notation, a superscript
$st$ to refer to this stationary regime. Recalling that we defined
the effective temperature $\Te$  by the condition $\langle
x^2\rangle=1$ at the lowest order in $\e$, we immediately find
from Eq. (\ref{admomentvgl1}) that at order $\e$ the piston will
indeed develop a nonzero average systematic speed equal to $\e
g(0)/[2 f(1/2)]$.  The explicit value of $\Te$ follows from Eq.
(\ref{admomentvgl2}), implying at lowest order in $\e$ that
$\langle x^2\rangle=f(3/2)/f(1/2)=1$. In original variables, cf.
Eq. (\ref{defx}), these results read as follows:
\begin{eqnarray}\label{adTeff}
\Te=\sqrt{T_1 T_2}
\end{eqnarray}
and
\begin{eqnarray}\label{adSpeedLowest}
\Vgems=\frac{\sqrt{2\pi}}{4}\sqrt{\frac{m}{M}}
\left(\sqrt{\frac{k_B T_1}{M}}-\sqrt{\frac{k_B
T_2}{M}}\right)+\ldots.
\end{eqnarray}
Although there is no macroscopic force present (pressures on both
sides of the piston are equal), the piston attains a stationary
state with a non-zero average velocity toward the higher
temperature region. Fluctuations conspire with the spatial
asymmetry to induce a net motion in the absence of macroscopic
forces. It is also clear from Eq.(\ref{adSpeedLowest}) that the
net motion vanishes when $T_1=T_2$ and also in the macroscopic
limit $M\rightarrow\infty$. The above result was already derived
in \cite{gruber}, but the calculation presented here is
streamlined so as to allow for  a swift calculation of higher
order corrections.
\newline
At each of the next orders, a coupling arises to a next higher
order moment. We shall present here the results up to order
$\e^5$, requiring the evaluation of the moments $\langle x^2
\rangle$, $\langle x^3 \rangle$, $\langle x^4 \rangle$, $\langle
x^5\rangle$ and $\langle x^6 \rangle$, up to orders $\e^4$,
$\e^3$, $\e^2$, $\e^1$ and $\e^0$, respectively (cf. appendix A
for details of the calculation). The resulting expression for the
average stationary speed in the original variable $V$ up to fifth
order in $\e$ is:
\begin{eqnarray}\label{adV5}
\langle V\rangle&=&\left(\frac{m}{M}\right)^{1/2}\sqrt{\frac{\pi
k_B}{2M}}\frac{1}{2}
\left(\sqrt{T_1}-\sqrt{T_2}\right)\nonumber\\
&+&\left(\frac{m}{M}\right)^{3/2}\sqrt{\frac{\pi k_B}{2M}}\left(
\frac{1}{4}\left(\sqrt{T_1}-\sqrt{T_2}\right)
-\frac{1}{3}\frac{\rho_1\sqrt{T_2}+\rho_2
\sqrt{T_1}}{\rho_1\sqrt{T_1}+\rho_2\sqrt{T_2}}\left(\sqrt{T_1}-\sqrt{T_2}\right)
%
+\frac{\pi}{16}\frac{\left(\sqrt{T_1}-\sqrt{T_2}\right)^3}
{\sqrt{T_1T_2}}
\right)\nonumber\\
%
&+&\left(\frac{m}{M}\right)^{5/2}\sqrt{\frac{\pi
k_B}{2M}}\frac{1}{8}\left( \left(\sqrt{T_1}-\sqrt{T_2}\right)
-5\frac{\left(\rho_1 T_1^{-1/2}+\rho_2 T_2^{-1/2}\right)^2
T_1T_2\left(\sqrt{T_1}-\sqrt{T_2}\right)}{\left(\rho_1
\sqrt{T_1}+\rho_2 \sqrt{T_2}\right)^2}\right.
\nonumber\\
%
 &
 &
+\frac{85}{18}\frac{\rho_1\sqrt{T_2}+\rho_2
\sqrt{T_1}}{\rho_1\sqrt{T_1}+\rho_2\sqrt{T_2}}\left(\sqrt{T_1}-\sqrt{T_2}\right)
+ \frac{1}{3}\frac{\rho_1 T_1^{-3/2}+\rho_2
T_2^{-3/2}}{\rho_1\sqrt{T_1}+\rho_2\sqrt{T_2}}T_1T_2\left(\sqrt{T_1}-\sqrt{T_2}\right)
-\frac{29\pi}{12}\frac{\left(\sqrt{T_1}-\sqrt{T_2}\right)^3}{\sqrt{T_1T_2}}\nonumber\\
%
%
%
%
& & +\frac{47\pi}{12}\frac{\left(\rho_1 T_1^{-1/2}+\rho_2
T_2^{-1/2}\right)
\left(\sqrt{T_1}-\sqrt{T_2}\right)^3}{\left(\rho_1
\sqrt{T_1}+\rho_2 \sqrt{T_2}\right)}
-\left.\frac{3\pi^2}{4}\frac{\left(\sqrt{T_1}-\sqrt{T_2}\right)^5}{T_1T_2}
\right)+\ldots.
\end{eqnarray}
As required, the average speed is zero at equilibrium, when
$T_1=T_2$. Note also that the average speed depends on the
densities of the gases solely through their ratio $\rho_1/\rho_2$.
This implies that, for $T_1$ and $T_2$ fixed, varying the
densities will not modify the steady-state velocity when operating
at mechanical equilibrium. In Fig. \ref{fig:speedAd}a. and Fig.
\ref{fig:speedAd}b., we illustrate the dependence of $\langle
V\rangle$ on the temperatures: the piston always moves towards the
high temperature region and its speed increases with the
temperature difference.

\section{Thermal Brownian motor}\label{s3}

\subsection{Motivation}

The systematic motion observed in the adiabatic piston is not
entirely surprising since the piston is embedded in a
nonequilibrium state with an explicit spatial asymmetry of its
surroundings. More interesting is the case of the thermal Brownian
motor, which was introduced and studied by molecular dynamics in a
recent paper \cite{vandenbroeckprl}. While the spatial environment
is perfectly symmetric here, the object itself has a spatial
asymmetry. The nonequilibrium conditions are generated by its
interaction with two (or more) ideal gases that are not at the
same temperature. The perturbative analysis, presented for the
adiabatic piston, can be repeated here but is more involved
because the problem is now genuinely two-dimensional.

\subsection{Presentation of the model}

\begin{figure}
\begin{center}
\includegraphics[width=5cm]{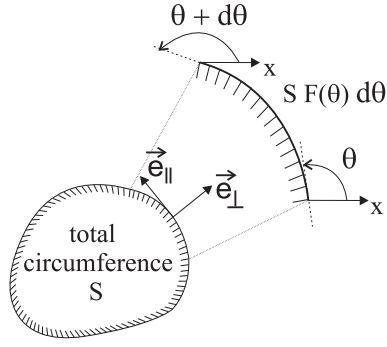}
\caption{A closed and convex object with total circumference $S$.
The length of the surface with an orientation between $\theta$ and
$\theta+d\theta$ is $SF(\theta)d\theta$, defining the form factor
$F(\theta)$.} \label{fig:Ftheta}
\end{center}
\end{figure}

Consider a $2$-dimensional convex and closed object with total
circumference $S$. Suppose that $dS$ is a small part of the
surface, inclined at an angle $\theta$, measured counterclockwise
from the $x$-axis (see Fig. \ref{fig:Ftheta}). We define the form
factor $F(\theta)$ as the fraction of the surface with orientation
$\theta$. This means that $S F(\theta)d\theta$ is the length of
the surface with orientation between $\theta$ and
$\theta+d\theta$. One can immediately verify that $F$ satisfies
\begin{eqnarray}\label{eigF}
\left\{\begin{array}{lll}
F(\theta)\geqslant 0 & \text{positivity} & \text{\ \ \ \ (a)}\\
\int_{0}^{2\pi}{d\theta F(\theta)=1} & \text{normalization}& \text{\ \ \ \ (b)}\\
\int_{0}^{2\pi}{d\theta F(\theta) \si}=\int_{0}^{2\pi}{d\theta
F(\theta)\co }=0 & \text{object is closed.}& \text{\ \ \ \ (c)}
\end{array}\right.
\end{eqnarray}
To simplify notation we will write $\sigem$ instead of
$\int_0^{2\pi}{d\theta F(\theta)\si}$.

We suppose that the object, with total mass $M$ and velocity
$\vec{V}$, has no rotational degree of freedom and a single
translational degree of freedom. Choosing the latter oriented
following the x-axis, we can write $\vec{V}=(V,0)$. Collisions of
the gas particles, of mass $m$ and velocity $\vec{v}$, with the
object are supposed to be instantaneous and perfectly elastical.
Hence, pre-collisional and post-collisional velocities of the
object, $V'$ and $V$, and of a gas particle,
$\vec{v'}=(v_x',v_y')$ and $\vec{v}=(v_x,v_y)$, are linked by
conservation of the total energy and the momentum in the
$x$-direction,
\begin{eqnarray}
 \frac{1}{2}M V'^2 +\frac{1}{2}m v_x'^2 +
\frac{1}{2}m v_y'^2 &=& \frac{1}{2}M V^2 +\frac{1}{2}m v_x^2 +
\frac{1}{2}m v_y^2\label{consenerg}\\
mv_x'+MV'&=&mv_x+MV. \label{xmoment}
\end{eqnarray}
Furthermore, we assume a (short-range) central force, implying
that the component of the momentum of the gas particle along the
contact surface of the object is conserved:
\begin{eqnarray}\label{centralforce}
\vec{v}'\cdot \vec{e}_{\parallel}=\vec{v}\cdot
\vec{e}_{\parallel},
\end{eqnarray}
with $\vec{e}_{\parallel}=(\co,\si)$, see Fig. \ref{fig:Ftheta}.
This yields for the post-collisional speed $V$:
\begin{eqnarray}\label{Btheta}
V=V' + \frac{2\frac{m}{M}\sit}{1+\frac{m}{M}\sit}\left(v_x'-
V'-v_y'\cotg \right).
\end{eqnarray}

\begin{figure}
\begin{center}
\includegraphics[width=6cm]{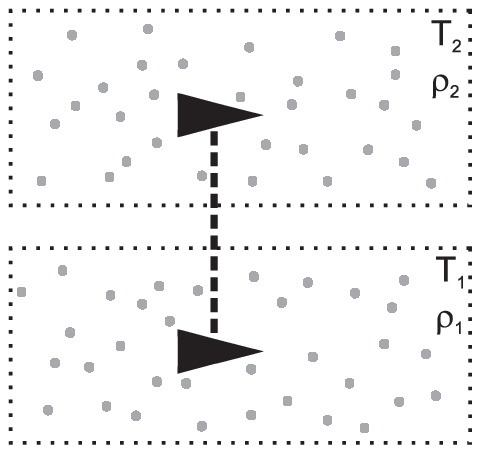}
\hspace{1cm}
\includegraphics[width=6cm]{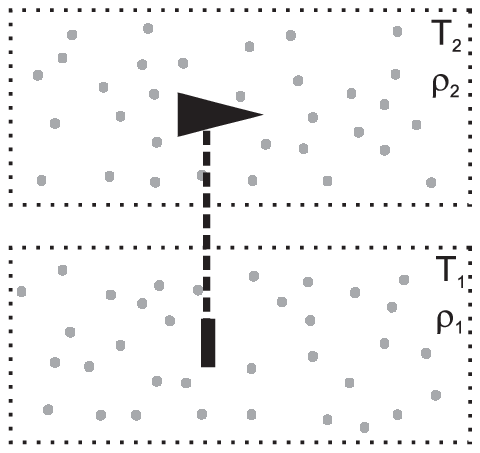}\\
(a)\hspace{7cm}(b)\\
\vspace{1cm}
\includegraphics[width=6cm]{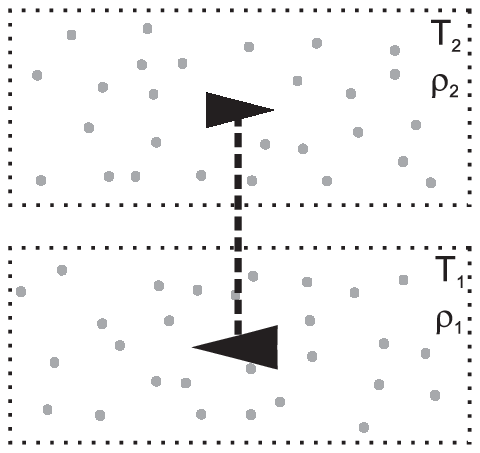}
\hspace{1cm}
\includegraphics[width=6cm]{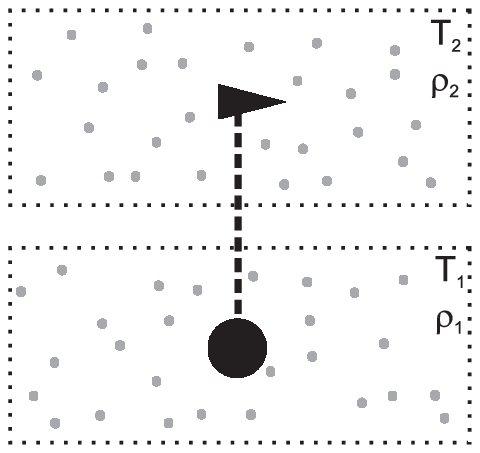}\\
(c)\hspace{7cm}(d)\\
\caption{Four different realizations of the thermal Brownian
motor, each consisting of two rigidly linked units. The units are
simple convex objects: a bar of length $L$, a disk of radius $R$
or an isosceles triangle with apex angle $2\theta_0$ and base $L$.
(a) was introduced in \cite{vandenbroeckprl} and will be referred
to here as \emph{Triangula}.} \label{fig:models}
\end{center}
\end{figure}

Similar as in the adiabatic piston problem, we start from the
linear Boltzmann equation (\ref{mastereq}), which is exact in the
ideal gas limit, to describe the motion of the object. The object
consists out of rigidly linked (closed and convex) parts, each
sitting in a reservoir $i$ containing an ideal gas with uniform
particle density $\rho_i$ and Maxwellian velocity distribution
$\phi_i$ at temperature $T_i$:
\begin{eqnarray}
\phi_i(v_x,v_y)=\frac{m}{2 \pi k_B
T_i}\exp{\left(\frac{-m(v_x^2+v_y^2)}{2 k_B T_i}\right)}.
\end{eqnarray}
Examples of the construction with two reservoirs are schematically
represented in Fig. \ref{fig:models}. The transition probability
$W(V|V')$ is then the sum of the contributions of the different
units of the object and can be calculated, starting from the basic
arguments of the kinetic theory. The contribution $dW_i$ to
$W(V|V')$ of the surface section of size $dS_i$, with orientation
in $[\theta,\theta+d\theta]$, exposed to the gas mixture $i$ is
\begin{eqnarray}
dW_i(V|V')&=&S_iF_i(\theta)d\theta\int_{-\infty}^{+\infty}{dv_x'\int_{-\infty}^{+\infty}{dv_y'}}
H\left[(\vec{V}'-\vec{v}')\cdot\vec{e}_{\bot}\right]
\left|(\vec{V}'-\vec{v}')\cdot\vec{e}_{\bot}\right| \nonumber\\
& &
\rho_i\phi_i(v_x',v_y')\delta\left[V-V'-\frac{2\frac{m}{M}\sit}{1+\frac{m}{M}\sit}
(v_x'-V'-v_y'\cotg)\right],\label{tranprobinfinites}
\end{eqnarray}
with $H$ the Heaviside function, $\delta$ the Dirac distribution
and $\vec{e}_{\bot}=(\si,-\co)$ a unit vector normal to the
surface, see Fig. \ref{fig:Ftheta}. The total transition
probability is then given by
\begin{eqnarray}
W(V|V')=\sum_i\int_0^{2\pi}{dW_i(V|V').}\label{transprob}
\end{eqnarray}
The integrals over the speed of the colliding gas particles can be
performed explicitly, resulting in:
\begin{eqnarray}\label{gentransprob2}
W(V|V')&=&\frac{1}{4}\sum_i S_i\rho_i \sqrt{\frac{m}{2\pi k_B
T_i}}
\left((V'-V)H[V'-V]\int_{\si>0}+(V-V')H[V-V']\int_{\si<0}\right)\nonumber\\
& & d\theta F_i(\theta) \left(\frac{M}{m\si}+\si\right)^2
\exp{\left[-\frac{m
\left(V'+\frac{1}{2}\left[(V-V')\left(1+\frac{M}{m
\sit}\right)\right]\right)^2\sit}{2k_B T_i}\right]}.
\end{eqnarray}
The remaining integral depends on the shape of the object.

Following the general set-up, we switch to the dimensionless
variable $x=\sqrt{\frac{M}{k_B\Te}}V$, cf. (\ref{defx}), with the
effective temperature $\Te$ to be determined from the condition
$\langle x^2\rangle=1$. The exact expression for the $n$-th
rescaled jump moment $A_n(x)$, cf. (\ref{defAn}), is then:
\begin{eqnarray}\label{genAn}
A_n(x)&=&(-1)^n 2^{(3n-1)/2}\su \sqrt{\frac{k_B T_i}{\pi m}}
\left(\frac{T_i}{\Te}\right)^{n/2}\e^{-n/2}\nonumber\\
& &\left(\int_{\si<0}+(-1)^n \int_{\si>0}\right)d\theta
F_i(\theta)
\exp\left[-\frac{\Te \sit}{2 T_i}x^2\e^2\right]\left(\frac{\e^2\si}{1+\e^2\sit}\right)^n\nonumber\\
& & \left(\Gamma\left[1+\frac{n}{2}\right]\Phi\left[1+\frac{n}{2},
\frac{1}{2},\frac{\Te  \sit}{2 T_i}x^2\e^2\right]+\e
x\si\sqrt{\frac{2\Te}{T_i}}
\Gamma\left[\frac{3+n}{2}\right]\Phi\left[\frac{3+n}{2},
\frac{3}{2},\frac{\Te\sit}{2 T_i}x^2\e^2\right]\right)
\end{eqnarray}
with $\Gamma$ the Gamma function (\ref{gamma}) and $\Phi$ the
Kummer function (\ref{kummer}).

\subsection{Stationary speed}

The equations of moments (\ref{momentvglx}), together with the
explicit expressions for the jump moments (\ref{genAn}), provide
the starting point for a straightforward series expansion in terms
of the small parameter $\e$. The equations for the first and
second moment, expanded up to order $\e^5$ and $\e^4$
respectively, are given by ($\tau=\e^2 t$):
\begin{eqnarray}
\partial_{\tau} \xgem &=&\su
\sqrt{\frac{k_B T_i}{m}} \left[-2\sqrt{\frac{2}{\pi}} \sitgemi
\xgem
+\left(\sqrt{\frac{T_i}{\Te}}-\sqrt{\frac{\Te}{T_i}}\xtgem\right)\sidgemi\
\e
 \right.\nonumber\\
& & +\frac{1}{3}\sqrt{\frac{2}{\pi}}\left(6 \xgem -
\frac{\Te}{T_i}\xdgem\right)\siviergemi\ \e^2
-\left(\sqrt{\frac{T_i}{\Te}}-\sqrt{\frac{\Te}{T_i}}\xtgem\right)\sivijfgemi\ \e^3\nonumber\\
&
&+\left.\sqrt{\frac{2}{\pi}}\left(-2\xgem+\frac{1}{3}\frac{\Te}{T_i}\xdgem+
\frac{1}{60}\left(\frac{\Te}{T_i}\right)^2\xvijfgem\right)\sizesgemi\
\e^4+
\left(\sqrt{\frac{T_i}{\Te}}-\sqrt{\frac{\Te}{T_i}}\xtgem\right)\sizevengemi\
\e^5\right]
\nonumber\\
& &
+O(\e^6) \label{genvgl1stemomentx}\\
\partial_{\tau} \xtgem &=&\su\sqrt{\frac{k_B T_i}{m}}
\left[-4\sqrt{\frac{2}{\pi}}
\left(-\frac{T_i}{\Te}+\xtgem\right)\sitgem_{i}\right.
+2\left(4\sqrt{\frac{T_i}{\Te}}\xgem-\sqrt{\frac{\Te}{T_i}}\xdgem\right)
\sidgemi\ \e\nonumber\\
& &+
2\sqrt{\frac{2}{\pi}}\left(-4\frac{T_i}{\Te}+5\xtgem-\frac{1}{3}\frac{\Te}{T_i}\xviergem
\right)\siviergemi\ \e^2 +
2\left(-7\sqrt{\frac{T_i}{\Te}}\xgem+2\sqrt{\frac{\Te}{T_i}}\xdgem\right)\sivijfgemi\ \e^3\nonumber\\
&
&+\left.\sqrt{\frac{2}{\pi}}\left(-16\xtgem+\frac{7}{6}\frac{\Te}{T_i}\xviergem+
\frac{1}{30}\left(\frac{\Te}{T_i}\right)^2\xzesgem
\right)\sizesgemi\ \e^4\right]+O(\e^5). \label{genvgl2demomentx}
\end{eqnarray}

Note that a term of order $\e^{-1}$ in the series for
$\partial_{\tau}\xgem$ is zero because of the property
(\ref{eigF}.c). Such a term would correspond to a constant,
velocity-independent, force acting on the object. It should indeed
be zero, since each gas mixture separately is in equilibrium. From
Eq. (\ref{genvgl1stemomentx}) we also immediately recognize to
lowest order in $\e$ the linear relaxation law, written in
original variables as $M\partial_t\Vgem=-\gamma V$, with
$\gamma=\sum_i\gamma_i$ and $\gamma_i$ the linear friction
coefficient, due to the section of the motor sitting in gas
mixture $i$:
\begin{eqnarray}
\gamma_i=4S_i\rho_i\sqrt{\frac{k_BT_im}{2\pi}}\int_0^{2\pi}
{d\theta F_i(\theta)\sit}.\label{gammageneral}
\end{eqnarray}
At this level of the perturbation, the speed of the object is
zero: no rectification takes place at the level of linear
response.

In order to find the first non-zero contribution to the velocity
$\xgems$, we need the terms up to order $\e$. From the definition
of $\Te$, by the condition $\xtgems=1$ up to lowest order in $\e$,
we find from Eq. (\ref{genvgl2demomentx}):
\begin{eqnarray}
\Te=\frac{\sum_i{\gamma_i T_i}}{\sum_i{\gamma_i}}.
\end{eqnarray}
The lowest non-zero term for the average velocity then follows
from Eq. (\ref{genvgl1stemomentx}) and reads in the original
variable $V$:
\begin{eqnarray}
\Vgems=\sqrt{\frac{m}{M}}\sqrt{\frac{\pi k_B\Te}{8M}} \frac{\su
\left(\frac{T_i}{\Te} - 1\right) \int_0^{2\pi}d\theta F_i (\theta)
\sin^3 \theta}{\su \sqrt{\frac{T_i}{\Te}} \int_0^{2\pi} d\theta
F_i(\theta) \sin^2(\theta)}+\ldots.\label{GenV1}
\end{eqnarray}
This speed is equal to the expansion parameter times the thermal
speed of the motor and further multiplied by a factor that depends
on the geometric properties of the object. Note that the Brownian
motor ceases to function in the absence of a temperature
difference (when $T_i = T_{\text{eff}}$ for all $i$) and in the
macroscopic limit $M\rightarrow \infty$ (since $\langle V \rangle
\sim 1/M$). Note also that the speed is scale-independent, i.e.,
independent of the actual size of the motor units: $\langle V
\rangle$ is invariant under the rescaling $S_i$ to $C S_i$.  To
isolate more clearly the effect of the asymmetry of the motor on
its speed, we focus on the case where the units have the same
shape in all compartments, i.e. $F_i(\theta)\equiv F(\theta)$ and
$S_i\equiv S$. One finds:
\begin{eqnarray}
\Te=\frac{\sum_i \rho_i T_i^{3/2}}{\sum_i\rho_i\sqrt{T_i}},
\end{eqnarray}
and
\begin{eqnarray}
\Vgems=\sqrt{\frac{m}{M}}\sqrt{\frac{\pi k_B\Te}{8 M}}
\frac{\sum_i \rho_i \left(\frac{T_i}{\Te}-1\right)}{\sum_i \rho_i
\sqrt{\frac{T_i}{\Te}}} \frac{\sidgem}{\sitgem}+\ldots.
\label{GenV1EqShape}
\end{eqnarray}
In this case $T_{\text{eff}}$ is independent of $ F(\theta)$ and
the drift velocity is proportional to $\langle
\sin^3\theta\rangle/\langle \sin^2\theta\rangle$, with the average
defined with respect to $F(\theta)$.  The latter ratio is in
absolute value always smaller than $1$, a value that can be
reached for ``strongly'' asymmetric objects as will be shown below
on specific examples. The resulting speed is then very large,
i.e., comparable to the thermal speed.

Calculation of the average speed up to order $\e^5$ requires the
evaluation of $\xtgem$, $\xdgem$, $\xviergem$, $\xvijfgem$ and
$\xzesgem$, up to orders $\e^4$, $\e^3$, $\e^2$, $\e$ and $\e^0$
respectively, cf. appendix B. This calculation has been carried
out using symbolic manipulations, and the resulting expressions
have been used in Table \ref{tab:tri} and Fig. \ref{fig:speedTri}.
However, since the analytic expressions are very involved, we only
reproduce the result here up to order $\e^3$:
\begin{eqnarray}\label{genV3}
\Vgems&=&\sqrt{\frac{m}{M}}\sqrt{ \frac{\pi k_B\Te}{8M}} \frac{\su
\left(\frac{T_i}{\Te} - 1\right) \sidgemi}{\su
\sqrt{\frac{T_i}{\Te}} \sitgemi}
\nonumber\\
&+&\left(\frac{m}{M}\right)^{3/2}\sqrt{\frac{\pi k_B
\Te}{8M}}\left\{ 
\frac{\su \left(1-\frac{T_i}{\Te}\right)\sivijfgemi}
{\su\sqrt{\frac{T_i}{\Te}}\sitgemi}\right.\nonumber\\
& &+\frac{\left(\su\sqrt{\frac{T_i}{\Te}}\siviergemi
\right)\left(\sum_i S_i\rho_i \left(\frac{T_i}{\Te}-\frac{7}{2}
\right)\sidgem_{_{i}}\right)}{\left[\su
\sqrt{\frac{T_i}{\Te}}\sitgem_{_{i}}\right]^2}\nonumber\\
& & +\frac{\su\sidgemi} {\su
\sqrt{\frac{T_i}{\Te}}\sitgemi}
\left[\frac{\su
\left(2\left(\frac{T_i}{\Te}\right)^{3/2}+\frac{1}{2}\sqrt{\frac{\Te}{T_i}}\right)\siviergemi}
{\su\sqrt{\frac{T_i}{\Te}}\sitgemi}\right.\nonumber\\
& &\ \ \ \ \ \ \ -\left.\frac{\pi}{2}\frac{\left(\su
\frac{T_i}{\Te}\sidgemi\right) \left(\su\left(\frac{T_i}{\Te}-1
\right)\sidgem_{_{i}}\right)}{\left[\su
\sqrt{\frac{T_i}{\Te}}\sitgem_{_{i}}\right]^2}
\right]\nonumber\\
& &+\left[\frac{\pi}{4}\left(\frac{\su \sidgemi} {\su
\sqrt{\frac{T_i}{\Te}}\sitgemi}\right)^2-\frac{1}{3}
\frac{\su\sqrt{\frac{\Te}{T_i}}\siviergemi}{\su
\sqrt{\frac{T_i}{\Te}}\sitgemi}\right]
\left[
\left.\frac{\su\left(-2
\left(\frac{T_i}{\Te}\right)^2+\frac{9}{2}\frac{T_i}{\Te}-\frac{5}{2}\right)\sidgemi}{\su
\sqrt{\frac{T_i}{\Te}}\sitgemi}
\right]\right\}\nonumber\\ &+&\ldots.\nonumber\\
\end{eqnarray}

\subsection{Special cases}

\begin{table}
\centering
\begin{tabular}{|c|c|c|c|}
\hline
 \textrm{Shape} & \ \ \ \textrm{Circumference $S$}\ \ \   & \textrm{Form factor F($\theta$) } &
  \textrm{Friction coefficient $\gamma$}\\
 \hline
 & & & \\
\textrm{Bar}   & $2L$ &
$\frac{1}{2}\left(\delta\left[\theta-\frac{\pi}{2}\right]+
\delta\left[\theta-\frac{3\pi}{2}\right]\right)$ & $8L\rho
\sqrt{\frac{k_B T m }{2\pi}}$
\\
 & & & \\
\textrm{Disk}   & $2\pi R$ & $1/2\pi$ & $4\pi R\rho\sqrt{\frac{k_B
T m}{2\pi}}$
\\
 & & & \\
\ \ \ \textrm{Triangle}\ \ \    & $L\frac{1+\sinul}{\sinul}$ & \ \
\
$\frac{2\delta\left[\theta-\frac{3\pi}{2}\right]\sinul+\delta\left[\theta-\theta_0\right]+
\delta\left[\theta-(\pi-\theta_0)\right]}{2(1+\sinul)}$\ \ \  & \
\ \ $4L\rho \sqrt{\frac{k_B T m}{2\pi}}(1+\sinul)$\ \ \
\\
 & & & \\
 \hline
\end{tabular}
\caption{The circumference $S$, the form factor $F(\theta)$ and
the friction coefficient $\gamma$ in a gas with density $\rho$ and
temperature $T$, for a vertical bar of length $L$, a disk with
radius $R$ and an isosceles triangle with base $L$ and apex angle
$2\theta_0$, pointed in the $x$-direction.} \label{tab:shapes}
\end{table}

\begin{table}
\centering
\begin{tabular}{|c|c|c|}
\hline \textrm{Shape} & \textrm{Figure}  & \textrm{Stationary
velocity $\Vgem$ (order $\e$)}
 \\
 \hline
 & &  \\
\textrm{Triangula}   & \ \ \ Fig. \ref{fig:models}a. \ \ \ &
$\frac{\sqrt{2 \pi k_B m}}{4M}(1-\sinul)
\frac{\rho_1\rho_2(T_1-T_2)(\sqrt{T_1}-\sqrt{T_2})}
{\left[\rho_1\sqrt{T_1}+\rho_2\sqrt{T_2}\right]^2}$
\\ & & \\
\textrm{Triangle - bar}   &  Fig. \ref{fig:models}b.   &
$\frac{\sqrt{2 \pi k_B
m}}{4M}(1-\sitnul)\frac{2\rho_1\rho_2\sqrt{T_1}(T_1-T_2)}{\left[2\rho_1
\sqrt{T_1}+\rho_2 \sqrt{T_2}(1+\sinul)\right]^2}$
\\ & & \\
\textrm{\ \ \ Triangle - triangle\ \ \ }   & Fig.
\ref{fig:models}c. & $\frac{\sqrt{2 \pi k_B m}}{4M}(1-\sinul)
\frac{\rho_1\rho_2(T_1-T_2)(\sqrt{T_1}+\sqrt{T_2})}
{\left[\rho_1\sqrt{T_1}+\rho_2\sqrt{T_2}\right]^2}$
\\ & & \\
\textrm{Triangle - disk}   &  Fig. \ref{fig:models}d.   & \ \ \
$\frac{\sqrt{2 \pi k_B m}}{4M}(1-\sitnul)\frac{\pi R}{L}
\frac{\rho_1\rho_2\sqrt{T_1}(T_1-T_2)}
{\left[\frac{\pi R}{L}\rho_1\sqrt{T_1}+(1+\sinul)\rho_2\sqrt{T_2}\right]^2}\ \ \ $\\
 & & \\
 \hline
\end{tabular}
\caption{Analytic result for the lowest order contribution to
$\Vgem$ for the different constructions depicted in Fig.
\ref{fig:models}.} \label{tab:models}
\end{table}

The above analytic result, Eq. (\ref{genV3}),  is valid for any
convex shape of the constituting pieces. To illustrate the type of
explicit results that are obtained, we focus on simple shapes like
a disk, a bar and a triangle. In Table \ref{tab:shapes} the
circumference $S$, the form factor $F(\theta)$ and the friction
coefficient $\gamma$ are calculated for these objects. Note that
the friction coefficient of the bar is in agreement with the
result of the adiabatic piston problem, cf. Eq. (\ref{adgamma}).

\begin{figure}
\begin{center}
\includegraphics[width=7cm]{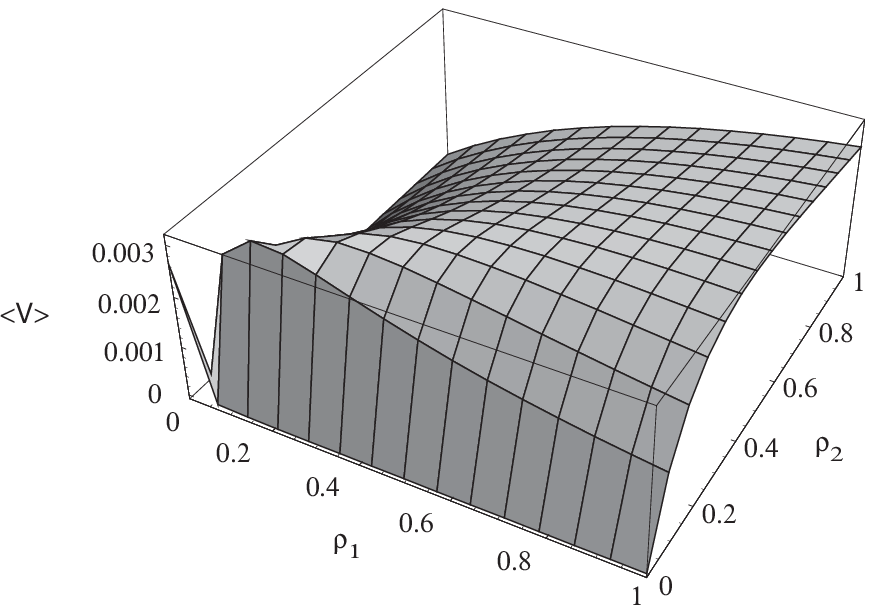}\hspace{2cm}
\includegraphics[width=7cm]{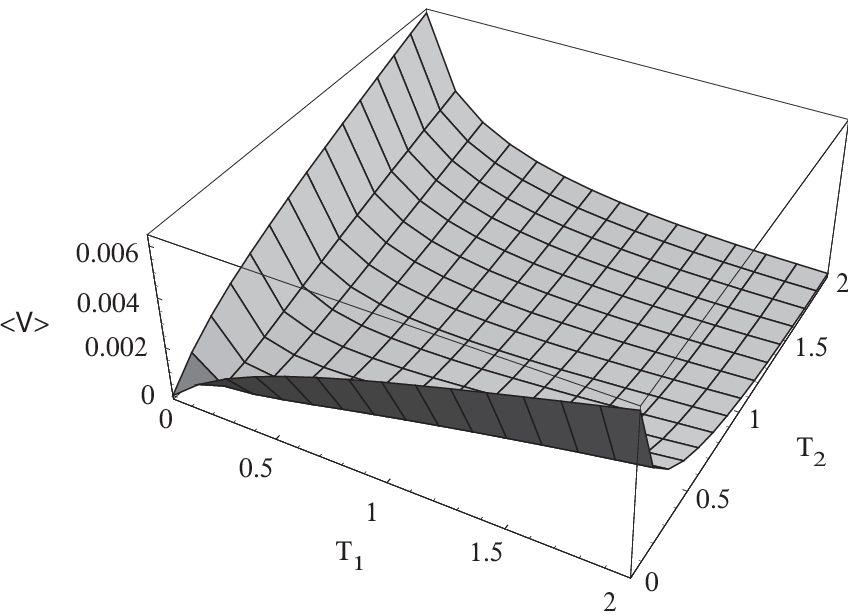}
\end{center}
\caption{Average speed of Triangula according to Eq. (\ref{Tri3}).
Left: the dependence of the velocity on the densities ($T_1=1.0$
and $T_2=5.0$). Right: The stationary speed increases with the
temperature difference ($\rho_1=\rho_2=0.0022$). The following
parameter values were used: mass of the gas particles $m=1$, total
mass of the motor $M=100$, apex angle of the triangles
$2\theta_0=\pi/18$ and $k_B=1$ by choice of units.}
\label{fig:speedTri}
\end{figure}

A thermal Brownian motor can only operate under nonequilibrium
conditions, which can be achieved if at least two of such units
are each located in a reservoir containing an ideal gas at a
different temperature. The two units are rigidly linked and can
move as a single degree of freedom along the $x$-direction.
Besides the nonequilibrium constraint, a spatial asymmetry is also
required to yield a net motion. In particular, a construction with
only bars and/or disks will not generate any net motion. One of
the simplest motors one can imagine was introduced in
\cite{vandenbroeckprl} and will be referred to as
\emph{Triangula}, see figure \ref{fig:models}a. Two identical
rigidly linked isosceles triangles (with apex angle $2\theta_0$,
pointing in the $x$-direction) are each located in a reservoir
containing a gas separately in equilibrium at a different
temperature. The lowest order contribution to the average velocity
of Triangula follows from Eq. (\ref{GenV1}), cf. Table
\ref{tab:shapes}:
\begin{eqnarray}
\Vgems_{\text{Triangula}}&=&\frac{\sqrt{2 \pi k_B
m}}{4M}(1-\sinul)
\frac{\rho_1\rho_2(T_1-T_2)(\sqrt{T_1}-\sqrt{T_2})}
{\left[\rho_1\sqrt{T_1}+\rho_2\sqrt{T_2}\right]^2}.\label{velocTT0}
\end{eqnarray}
The fact that the combination of the asymmetry and the temperature
gradient is necessary to generate the systematic motion, is
contained in Eq. (\ref{velocTT0}). If either $T_1=T_2$ or
$\theta_0=\pi/2$ (the triangle becomes a bar and thus the
asymmetry disappears), the average velocity vanishes. The
corrections of order $\e^3$ to this formula can be found in
appendix B. The dependence of the speed on the temperatures and
densities are reproduced in Fig. \ref{fig:speedTri}. Note that
from Eq. (\ref{velocTT0}) may be concluded that the speed of
Triangula is maximal for $\rho_1 \sqrt{T_1}=\rho_2 \sqrt{T_2}$.
Fig. \ref{fig:models} also shows some other variations of the
thermal Brownian motor. Their drift speed up to lowest order in
$\e$ can be found in Table \ref{tab:models}. One can easily verify
that each thermal Brownian motor ceases to function when the
spatial asymmetry or the temperature difference vanish. In this
context it is sometimes stated that equilibrium is a point of flux
reversal. This is indeed the case for our microscopic model except
when there are special symmetries in the system. When the units in
the two reservoirs are not the same, the direction of the net
motion changes when the temperature difference changes sign. From
the models depicted in Fig. \ref{fig:models}, only Triangula keeps
its original direction of motion. In this latter case, the speed
exhibits a parabolic minimum as a function of the temperature,
with zero speed at equilibrium. The reason for this peculiar
behavior derives from the permutational symmetry of identical
units, implying that the speed must be invariant under the
interchange of $T_1,\rho_1$ with $T_2,\rho_2$. In particular it
must be an even function of $T_1-T_2$ when $\rho_1=\rho_2$.

\section{Comparison with simulations}\label{s4}

The analytic results for the adiabatic piston and the thermal
Brownian motor are compared with the results of the numerical
solution of the Boltzmann equation in Table \ref{tab:ad} and Table
\ref{tab:tri} respectively. To improve the precision of these
simulations, we used a special technique for solving a Master
equation, based on the introduction of a simple envelope process,
see appendix C for details. The agreement between theory and
simulations is very satisfactory, and only breaks down, as
expected, for small ratios $m/M$ where the perturbative result
becomes inaccurate. We have also included for completeness the
results obtained by molecular dynamics simulations, cf.
\cite{vandenbroeckprl} for more details. We used low densities for
the hard disks, $\rho=.0022$, a regime in which one comes close to
the properties of an ideal gas. Nevertheless one expects  strong
finite size effects, due to the fact that the reservoirs
containing the gases are not very large. To cite just one
important phenomenon, we note that the motion of the motor will
generate sound waves that will reimpact on it. Taking this into
account, the speeds observed in the molecular dynamics are in
reasonable agreement with the theoretical and numerical results
obtained from the Boltzmann equation. Notably the speed of the
thermal motor in the hard disk gases is systematically larger by
roughly $20$ to $40\%$, for reasons that are unclear to us.

\begin{table}
\centering
\begin{tabular}{|c|c|c|c|c|}
\hline
 \textrm{Mass} & \textrm{Theory } &
 \textrm{Theory } & \textrm{Theory }  & \textrm{Boltzmann}\\
 $M$ & \textrm{(order $\e$)} & \textrm{(order $\e+\e^3$)} & \textrm{(order $\e+\e^3+\e^5$)} & \textrm{equation}\\
 \hline
 1    & 5.64    & -7.82    & 20.44   & 1.411 \\
 5    & 1.13    & 0.590    & 0.8158  & 0.7289 \\
 20   & 0.282   &  0.2484  & 0.2519  & 0.2511 \\
 50   & 0.113   & 0.1074   & 0.1076  & 0.1076 \\
\ \ \  100\ \ \    &\ \ \  0.0564 \ \ \  &\ \ \   0.05505 \ \ \ &
\ \ \   0.05508\ \ \   &\ \ \   0.0554 \ \ \  \\
200   & 0.0282  & 0.02786  & 0.02787 & 0.0280 \\
 \hline
\end{tabular}
\caption{Stationary average speed of the \emph{adiabatic piston}
from the perturbative solution method up to order $\e$, $\e^3$ and
$\e^5$, compared with the result from a numerical solution of the
Boltzmann equation. The following parameter values were used:
particle densities $\rho_1=10^{-2}$ and $\rho_2=1$, temperatures
$T_1=100$ and $T_2=1$ and mass of the gas particles $m=1$. $k_B=1$
by choice of units.}\label{tab:ad}
\end{table}

\begin{table}
\centering
  \begin{tabular}{|c|c|c|c|c|c|}
  \hline
   Mass           & Theory    & Theory    & Theory & Boltzmann & Molecular \\
   $M$           &    (order $\e$)   & (order $\e+\e^3$) & (order $\e+\e^3 + \e^5$) & equation & Dynamics    \\   \hline
  1            & 0.38                  & -1.23       &  11.66            & 0.057     &0.12\\
  5            & 0.076                 & 0.01188     &  0.1150           & 0.0470    &0.064\\
  20           & 0.0190                & 0.01502     &  0.01663          & 0.0157   &0.024\\
  50           & 0.00762               & 0.006974    &  0.007077         & 0.0071   &0.0093\\
  100          & 0.00381               & 0.003648    &  0.003661         & 0.0035   &0.0043\\
\ \ \    200\ \ \            &\ \ \   0.00190  \ \ \
 &\ \ \   0.001864\ \ \      &\ \ \    0.001866\ \ \           &\ \ \   0.0017 \ \ \
 &\ \ \  0.0021\ \ \  \\
  \hline
  \end{tabular}
\caption{Stationary average speed of the thermal Brownian motor
\emph{Triangula} from the perturbative solution method up to order
$\e$, $\e^3$ and $\e^5$, compared with the result from a numerical
solution of the Boltzmann equation. The following parameter values
were used: particle densities $\rho_1=\rho_2=0.00222$,
temperatures $T_1=1.9$ and $T_2=0.1$, mass of the particles $m=1$
and apex angle of the triangle $2\theta_0=\pi/18$.
$k_B=1$ by choice of units. For comparison, we also include the
speed observed in molecular dynamics, see \cite{vandenbroeckprl}
for more details.}\label{tab:tri}
\end{table}

\section{Discussion}
The problem of the Maxwell demon has been haunting the imagination
and theoretical efforts of physicists for more than hundred years.
While there is a consensus that one cannot rectify fluctuations at
equilibrium, it is recomforting that one can construct microscopic
models, involving interactions with ideal gases only, for which
this thesis can be verified explicitly. The same models can be
used as test cases for another important field of interest, namely
the rectification of thermal fluctuations in nonequilibrium, also
referred to as Brownian motors. In this respect we claim that our
model is a genuine Brownian motor: the rectification appears at
the level of nonlinear response, where the usual separation
between systematic and noise terms, as made  explicit in a linear
Langevin equation, is no longer possible. Hence the operation of
our Brownian motor falls outside the scope of linear irreversible
thermodynamics. It belongs to the realm of microscopic theory in
which nonlinearity and noise form an intertwined part of the
microscopic dynamics. This is in contrast to most of the Brownian
motors discussed with mesoscopic theory. For example, the
prototype model of one class of Brownian motors referred to as
flashing ratchet can be described by diffusion in an  external
potential, a standard problem in linear irreversible
thermodynamics.

\appendix

\section{Details on the calculation for the adiabatic piston}
Expansion of the rescaled jump moments $A_n(x)$ (\ref{adAn}) in
the moment equations (\ref{momentvglx}) yields the following power
series expansion in $\e=\sqrt{m/M}$, with the first and second
moment expanded up to fifth and fourth order in $\e$ (cf. Eqs.
(\ref{admomentvgl1}-\ref{admomentvgl2})) respectively, and the
higher moments $\xdgem, \xviergem, \xvijfgem$ and $\xzesgem$ up to
$\e^3, \e^2, \e$ and $\e^0$ respectively:
\begin{eqnarray}\label{adDmomentvgl1full}
\partial_{\tau}\xdgem&=&6\left(2\fdrietw\xgem-\fhalf\xdgem\right)-3\left(4\gtwee
+\gnul\xviergem\right)\e
\nonumber\\
& &
+\left(-56\fdrietw\xgem+24\fhalf\xdgem-\fminhalf\xvijfgem\right)\e^2+9\left(4\gtwee+\gnul\xviergem\right)\e^3+O(\e^4)
\nonumber\\
%
\partial_{\tau}\xviergem&=&8\left(3\fdrietw\xtgem-\fhalf\xviergem\right)-4\left(
12\gtwee\xgem\
+\gnul\xvijfgem\right)\e\nonumber\\
& & +4\left(16\fvijftw-44\fdrietw\xtgem+
11\fhalf\xviergem-\frac{1}{3}\fminhalf\xzesgem\right)\e^2+O(\e^3)\nonumber\\
\partial_{\tau}\xvijfgem&=&10\left(4\fdrietw\xdgem-\fhalf\xvijfgem\right)
-5\left(24\gtwee\xtgem+\gnul\xzesgem\right)\e+O(\e^2)\nonumber\\
\partial_{\tau}\xzesgem&=&12\left(5\fdrietw\xviergem-\fhalf\xzesgem\right)+O(\e).
\end{eqnarray}
In the stationary regime these equations form, together with Eqs.
(\ref{admomentvgl1}-\ref{admomentvgl2}), an algebraic set of
equations, which can be solved to find the stationary average
velocity $\xgem$ up to order $\e^5$. The result in terms of the
original variable $V=\sqrt{\frac{k_B \Te}{M}} x$ is given in Eq.
(\ref{adV5}). The corresponding results for the higher order
moments read:
\begin{eqnarray}
\langle V^2\rangle&=&\frac{k_B \sqrt{T_1T_2}}{M}\nonumber\\
&+&\frac{k_B m}{M^2} \left(\frac{\sqrt{T_1T_2}}{2}
+\frac{\pi}{8}\left(\sqrt{T_1}-\sqrt{T_2}\right)^2
-\frac{1}{2}\frac{\left(\rho_1 T_1^{-1/2}+\rho_2
T_2^{-1/2}\right)T_1 T_2}{\rho_1 \sqrt{T_1}+\rho_2 \sqrt{T_2}}
\right)\nonumber\\
&+&\frac{k_B m^2}{M^3}\left(\frac{\sqrt{T_1T_2}}{4} -\frac{2}{3}
\frac{\left(\rho_1 T_1^{-1/2}+\rho_2
T_2^{-1/2}\right)^2(T_1T_2)^{3/2}}{\left(\rho_1 \sqrt{T_1}+\rho_2
\sqrt{T_2}\right)^2}\right.
\nonumber\\
& & -\frac{29\pi}{48} \left(\sqrt{T_1}-\sqrt{T_2}\right)^2
-\frac{3\pi^2}{16}
\frac{\left(\sqrt{T_1}-\sqrt{T_2}\right)^4}{\sqrt{T_1T_2}}
+\frac{35\pi}{48} \frac{\rho_1 T_1^{5/2}+\rho_2 T_2^{5/2}}{\rho_1
\sqrt{T_1}+\rho_2
\sqrt{T_2}}\frac{\left(\sqrt{T_1}-\sqrt{T_2}\right)^2}{T_1T_2}
\nonumber\\
& & +\left. \frac{1}{8}\frac{\left(\rho_1 T_1^{-3/2}+\rho_2
T_2^{-3/2}\right)(T_1T_2)^{3/2}}{\rho_1 \sqrt{T_1}+\rho_2
\sqrt{T_2}} +\frac{7}{24}\frac{\left(\rho_1 T_1^{-1/2}+\rho_2
T_2^{-1/2}\right)T_1T_2}{\rho_1 \sqrt{T_1}+\rho_2 \sqrt{T_2}}
\right)+\ldots\nonumber\\
\langle V^3\rangle&=&\frac{\sqrt{k_B^3 m \pi}}{2\sqrt{2}M^2}
\left(\sqrt{T_1}-\sqrt{T_2}\right)\sqrt{T_1T_2}
\nonumber\\
&+& \frac{\sqrt{k_B^3 m^3\pi}}{\sqrt{2}M^3} \left(
-\frac{8}{3}\left(\sqrt{T_1}-\sqrt{T_2}\right)\sqrt{T_1T_2}
 -\frac{3\pi}{4}\left(\sqrt{T_1}-\sqrt{T_2}\right)^3
+ \frac{7}{4}\frac{\left(\sqrt{T_1}-\sqrt{T_2}\right) \left(\rho_1
T_1^{5/2}+\rho_2 T_2^{5/2}\right)}{\left(\rho_1 \sqrt{T_1}+\rho_2
\sqrt{T_2}\right)\sqrt{T_1T_2}}
\right)+\ldots\nonumber\\
\langle V^4\rangle&=&3\left(\frac{k_B\sqrt{T_1T_2}}{M}\right)^2
+\frac{k_B^2 m}{M^3} \left(-4T_1T_2
-\frac{7\pi}{4}\left(\sqrt{T_1}-\sqrt{T_2}\right)^2\sqrt{T_1T_2}
+4\frac{\rho_1 T_1^{5/2}+\rho_2 T_2^{5/2}}{\rho_1
\sqrt{T_1}+\rho_2
\sqrt{T_2}}\right)+\ldots\nonumber\\
\langle V^5\rangle&=&-\frac{\sqrt{k_B^5 m\pi}}{M^3}
\frac{5}{2\sqrt{2}}\left(\sqrt{T_1}-\sqrt{T_2}\right)T_1T_2+\ldots\nonumber\\
\langle V^6\rangle&=&15\left(\frac{k_B
\sqrt{T_1T_2}}{M}\right)^3+\ldots.
\end{eqnarray}
Note that the lowest order terms are consistent with the
observation that the velocity distribution itself is Gaussian at
the lowest order, cf. \cite{gruber}.

\section{Details on the calculation for a thermal Brownian motor of a general shape}

A perturbative series in $\e$ for the moments, obtained by the
expansion of Eqs. (\ref{momentvglx}), together with the jump
moments defined in (\ref{adAn}). The resulting expressions for the
first and second moment up to order $\e^5$ and $\e^4$ respectively
are given in Eqs.
(\ref{genvgl1stemomentx}-\ref{genvgl2demomentx}). Calculation of
$\Vgem$ up to $\e^5$ requires furthermore $\xdgem$, $\xviergem$,
$\xvijfgem$ and $\xzesgem$ up to order $\e^3$, $\e^2$, $\e$ and
$\e^0$ respectively:
\begin{eqnarray}
\partial_{\tau}\xdgem &=&\su\sqrt{\frac{k_B
T_i}{m}}  \left\{6\sqrt{\frac{2}{\pi}}
\left(2\frac{T_i}{\Te}\xgem-\xdgem\right)\sitgemi\right.
\nonumber\\
& &+3\left.\left(-4\left(\frac{T_i}{\Te}\right)^{3/2}
+7\sqrt{\frac{T_i}{\Te}}\xtgem-\sqrt{\frac{\Te}{T_i}}\xviergem\right)
\sidgemi\ \e\right.\nonumber\\
& &
+\sqrt{\frac{2}{\pi}}\left(-56\frac{T_i}{\Te}\xgem+24\xdgem-\frac{\Te}{T_i}\xvijfgem\right)\siviergemi\
\e^2
\nonumber\\
& &
+\left.3\left(12\left(\frac{T_i}{\Te}\right)^{3/2}-21\sqrt{\frac{T_i}{\Te}}\xtgem+3\sqrt{\frac{\Te}{T_i}}\xviergem\right)
\sivijfgemi\ \e^3\right\}
+O(\e^4)\nonumber\\
\label{vgl3demomentx}
\partial_{\tau}\xviergem&=&4\su\sqrt{\frac{k_B
T_i}{m}} \left\{2\sqrt{\frac{2}{\pi}}
\left(3\frac{T_i}{\Te}\xtgem-\xviergem\right)\sitgemi\right.\nonumber\\
& &+\left(-12\left(\frac{T_i}{\Te}\right)^{3/2}\xgem+10
\sqrt{\frac{T_i}{\Te}}\xdgem-\sqrt{\frac{\Te}{T_i}}\xvijfgem\right)\sidgemi\ \e\nonumber\\
& &
+\left.\sqrt{\frac{2}{\pi}}\left(16\left(\frac{T_i}{\Te}\right)^2
-
44\frac{T_i}{\Te}\xtgem+11\xviergem-\frac{1}{3}\frac{\Te}{T_i}\xzesgem\right)\siviergemi\
\e^2\right\}
+O(\e^3)\nonumber\\
\partial_{\tau}\xvijfgem&=&5\su\sqrt{\frac{k_B
T_i}{m}}\left\{2\sqrt{\frac{2}{\pi}}\left( 4
\frac{T_i}{\Te}\xdgem-\xvijfgem\right)\sitgemi\right.\nonumber\\
& & +\left(\left.
-24\left(\frac{T_i}{\Te}\right)^{3/2}\xtgem+13\sqrt{\frac{T_i}{\Te}}\xviergem-
\sqrt{\frac{\Te}{T_i}}\xzesgem\right)\sidgemi\ \e\right\}+O(\e^2)\nonumber\\
\partial_{\tau}\xzesgem&=&12\su\sqrt{\frac{k_B
T_i}{m}}\left\{\sqrt{\frac{2}{\pi}}\left(5\frac{T_i}{\Te}\xviergem-\xzesgem\right)\sitgemi\right\}+O(\e)
\end{eqnarray}
In the stationary regime, these equations form together with Eqs.
(\ref{genvgl1stemomentx}-\ref{genvgl2demomentx}) an algebraic set
of equations from which the average velocity can be obtained up to
order $\e^5$. The analytical expression in the original variable
$V$ up to $\e^3$ is reproduced in Eq. (\ref{genV3}). The
corresponding power series for the higher moments of the
stationary velocity distribution function are:
\begin{eqnarray}
\langle V^2\rangle&=&\frac{k_B \Te}{M}+\frac{k_B m \Te}{M^2}
\left\{\frac{\su
\left(-2\left(\frac{T_i}{\Te}\right)^{3/2}+\frac{5}{2}\sqrt{\frac{T_i}{\Te}}-
\frac{1}{2}\sqrt{\frac{\Te}{T_i}}\right)\siviergemi} {\su
\sqrt{\frac{T_i}{\Te}}\sitgemi}
\right.\nonumber\\
& &-\left.\frac{\pi}{4}\frac{\left(\su \sidgemi\right)\left( \su
\left(-2
\left(\frac{T_i}{\Te}\right)^2+\frac{13}{2}\frac{T_i}{\Te}
-\frac{5}{2}\right)\sidgemi\right)-2\left(\su
\frac{T_i}{\Te}\sidgemi\right)^2}{\left[\su
\sqrt{\frac{T_i}{\Te}}\sitgemi\right]^2}\right\}+\ldots\nonumber\\
\langle V^3\rangle&=&\sqrt{\frac{m}{M}}\left(\frac{k_B
\Te}{M}\right)^{3/2}\left[
\sqrt{\frac{\pi}{2}}\frac{\su\left(-2
\left(\frac{T_i}{\Te}\right)^2+\frac{9}{2}\frac{T_i}{\Te}
-\frac{5}{2}\right)\sidgemi}{\su\sqrt{\frac{T_i}{\Te}}\sitgemi}\right]+\ldots\nonumber\\
\langle
V^4\rangle&=&3\left(\frac{k_B\Te}{M}\right)^2+\ldots\nonumber\\
\langle V^5\rangle&=&\sqrt{\frac{m}{M}}\left(\frac{k_B
\Te}{M}\right)^{5/2}\left[
\sqrt{\frac{\pi}{2}}\frac{\su
\left(-20\left(\frac{T_i}{\Te}\right)^2+\frac{75}{2}\frac{T_i}{\Te}
-\frac{35}{2}\right)\sidgemi} {\su
\sqrt{\frac{T_i}{\Te}}\sitgemi}\right]+\ldots\nonumber\\
\langle V^6\rangle&=&15\left(\frac{k_B\Te}{M}\right)^3+\ldots.
\end{eqnarray}

For the particular case of Triangula, see Fig. \ref{fig:models}a.,
one can verify that
\begin{eqnarray}
\Te&=&\frac{\rho_1 T_1^{3/2}+\rho_2
T_2^{3/2}}{\rho_1\sqrt{T_1}+\rho_2\sqrt{T_2}}\nonumber\\
\langle
\sin^n{\theta}\rangle&=&\frac{(-1)^n\sinul+\sin^n{\theta_0}}{1+\sinul}
\end{eqnarray}
The average speed of Triangula up to $\e^3$ reads:
\begin{eqnarray}\label{Tri3}
\Vgem&=&\sqrt{\frac{m}{M}}\sqrt{\frac{\pi
k_B\Te}{2M}}\frac{1}{2}\left(\heen-\hnul\right)\left(\sinul-1\right)\nonumber\\
& &\left(\frac{m}{M}\right)^{3/2}\sqrt{\frac{\pi
k_B\Te}{2M}}\left\{\frac{1}{2}\left(\hnul-\heen\right)\frac{\siviernul-1}{\sinul+1}
+\frac{\pi\hnul}{4}\left(-\heen^2
-\hnul\htwee+\frac{13}{4}\hnul\heen-\frac{5}{4}\hnul^2\right)\left(\sinul-1\right)^3\right.\nonumber\\
& & \ \ \ \ +
\left(\frac{1}{2}\hhalf\heen-\frac{7}{4}\hhalf\hnul+\hnul\hdrietw+
\frac{1}{3}\hminhalf\htwee-\frac{3}{4}\hminhalf\heen+\frac{2}{3}\hminhalf\hnul\right)\nonumber\\
& & \ \ \ \ \  \ \ \ \ \  \times\left.
\frac{\left(\sinul-1\right)\left(1+\sidnul\right)}{1+\sinul}\right\},
\end{eqnarray}
with
\begin{eqnarray}
h[n]=\frac{\rho_1 T_1^{n}+\rho_2
T_2^{n}}{\rho_1\sqrt{T_1}+\rho_2\sqrt{T_2}}\ \Te^{-n+1/2}.
\end{eqnarray}

\section{Simulation of the Master equation}

Consider a Markov process, defined by the transition rate
$W(V|V')$, for transitions from state $V'$ to $V$. We decompose
the transition rate as follows
\[
W(V|V') = R(V') P(V|V')
\]
where $R(V') = \int dV\,W(V|V')$ is the total rate and $P(V|V')$
is the conditional probability. Stochastic trajectories of $V$ may
be easily produced if the probability $P$ can be easily realized,
that is, if it is simple to generate random values with that
distribution. However, for complicated $W$ and $P$ this direct
approach is often not possible and an alternative construction is
necessary.

We introduce an ``envelope'' process, $W^*(V|V')$, for which $W^*
\geq W$ for all $V$,$V'$, cf. \cite{ethier}. The difference
\[
W_0(V')= W^*(V|V') - W(V|V')
\]
is called the ``null'' process. The envelope process is chosen
such that $R^*$ is a constant and that $P^*$ has a simple form
(e.g., a Gaussian distribution).

The simulation of trajectories of $W$ then proceeds as follows:

1) Given the current state $V'$, find the time to the next
transition for the envelope process as
\[
\tau = -\ln(\Re)/R^*
\]
where $\Re$ is a uniformly distributed random value in $(0,1)$.
The random variable $\tau$ is exponentially distributed with
$\langle \tau \rangle = 1/R^*$.

2) Choose the new state, $V$, from the distribution $P^*(V|V')$.

3) With probability $W(V|V')/W^*(V|V')$ the transition $V'
\rightarrow V$ occurs, otherwise, the event is a null event and
the state $V'$ is unchanged. In either case, the time is advanced
by $\tau$.

4) Return to step 1) until the required number of iterations are
performed.

Note that the probability of a transition $V' \rightarrow V$ being
selected in step 2) is $P^*(V|V')$ and the probability of that
transition being accepted in step 3) is $W/W^*$ so the net
probability of the accepted transition is
\[
P^*(V|V')\frac{W(V|V')}{W^*(V|V')} = P^*(V|V')\frac{W(V|V')}{R^*
P^*(V|V')} = \frac{1}{R^*} W(V|V')
\]
Since the total rate (accepted plus null transitions) is $R^*$ the
algorithm produces the stochastic process with the desired
transition rate $W(V|V')$.

Clearly, the method will be inefficient if the ratio $W/W^*$ is
small since most transitions would be rejected. On the other hand,
the method is only correct if  $W/W^* \leq 1$ for all $V,V'$ since
the probability of acceptance cannot be greater than one (i.e.,
the null process cannot have negative probability for any $V'$).
As such, the form of the envelope process (e.g., the mean and
variance of the Gaussian) must be chosen with care.

\end{document}